\documentclass[useAMS,usenatbib]{mn2e}

\input psfig.sty
\usepackage{graphicx}
\usepackage{amsmath}
\usepackage{xspace}
\usepackage{bbm}

\newcommand {\apj} {ApJ}

\newcommand {\etal} {et~al.~}
\def \spose#1{\hbox  to 0pt{#1\hss}}  
\newcommand {\lta} {\mathrel{\spose{\lower 3pt\hbox{$\sim$}}\raise  2.0pt\hbox{$<$}}}
\newcommand {\gta} {\mathrel{\spose{\lower  3pt\hbox{$\sim$}}\raise 2.0pt\hbox{$>$}}}

\newcommand {\ha}  {\ifmmode H\alpha \else H$\alpha $ \fi} 

\def\Sref#1{Section~\ref{#1}\xspace}
\def\Fref#1{Figure~\ref{#1}\xspace}
\def\Tref#1{Table~\ref{#1}\xspace}
\def\Eref#1{Equation~\ref{#1}\xspace}

\newcommand {\kms} {\ifmmode  \,\rm km\,s^{-1} \else $\,\rm km\,s^{-1}  $ \fi }
\newcommand {\kpc} {\ifmmode  {\rm kpc}  \else ${\rm  kpc}$ \fi  }  
\newcommand {\pc} {\ifmmode  {\rm pc}  \else ${\rm pc}$ \fi  }  
\newcommand {\Msun} {\ifmmode {\rm M_{\odot}} \else ${\rm M_{\odot}}$ \fi} 
\newcommand {\Zsun} {\ifmmode {\rm Z_{\odot}} \else ${\rm Z_{\odot}}$ \fi} 
\newcommand {\yr} {\ifmmode yr^{-1} \else $yr^{-1}$ \fi} 
\newcommand {\hMsun} {\ifmmode h^{-1}\,\rm M_{\odot} \else $h^{-1}\,\rm M_{\odot}$ \fi}

\def\sigmasie{\sigma_{\mathrm{SIE}}}

\def\q3{q_{3}}

\def\pr{{\rm Pr}}

\usepackage[usenames]{color}

\newcommand\alphabulge{\alpha_{\rm b}}
\newcommand\alphabulgemedian{\mu_{\alphabulge}}
\newcommand\alphabulgewidth{\sigma_{\alphabulge}}

\newcommand\alphadisc{\alpha_{\rm d}}
\newcommand\alphadiscmedian{\mu_{\alphadisc}}
\newcommand\alphadiscwidth{\sigma_{\alphadisc}}

\newcommand\SPSMbulge{\mathcal{M}_{\rm b}}
\newcommand\SPSMdisc{\mathcal{M}_{\rm d}}
\newcommand\SPSMbulgemedian{\mu_{\SPSMbulge}}
\newcommand\SPSMbulgewidth{\sigma_{\SPSMbulge}}
\newcommand\SPSMdiscmedian{\mu_{\SPSMdisc}}
\newcommand\SPSMdiscwidth{\sigma_{\SPSMdisc}}

\newcommand\mtot{m_{\rm tot}}
\newcommand\mbulge{m_{\rm b}}
\newcommand\mdisc{m_{\rm d}}
\newcommand\mhalo{m_{\rm h}}
\newcommand\Mtot{M_{\rm tot}}

\newcommand\fstar{f}
\newcommand\logfstar{F}

\newcommand\lfstar{\ell}
\newcommand\lfstargradient{\lfstar_1}
\newcommand\lfstarintercept{\lfstar_0}
\newcommand\lfstarwidth{\sigma_{\lfstar}}

\newcommand\obsSPSMbulge{\widetilde{\SPSMbulge}}
\newcommand\obsSPSMdisc{\widetilde{\SPSMdisc}}
\newcommand\obsMtot{\widetilde{\Mtot}}
\newcommand\obsSPSMbulgej{\widetilde{\SPSMbulge^j}}
\newcommand\obsSPSMdiscj{\widetilde{\SPSMdisc^j}}
\newcommand\obsMtotj{\widetilde{\Mtot^j}}
\newcommand\errSPSMbulge{\sigma_{\rm b}}
\newcommand\errSPSMdisc{\sigma_{\rm d}}
\newcommand\errMtot{\sigma_{\rm tot}}

\newcommand\jags{{\tt JAGS}}
\newcommand\daft{{\tt DAFT}}

\def\onegalaxy{SWELLS\ J0820$+$4847\xspace}

\def\ucsb{Dept. of Physics, University of California, 
  Santa Barbara, CA 93106, USA}
\def\kipac{Kavli Institute for Particle Astrophysics and Cosmology, 
 Stanford University, 452 Lomita Mall, Stanford, CA 94035, USA}

\def\oxford{Department of Physics, University of Oxford, 
  Keble Road, Oxford, OX1 3RH, UK}
\def\cambridge{Institute of Astronomy, University of Cambridge,
  Madingley Rd, Cambridge, CB3 0HA, UK}
\def\mpia{Max Planck Institue for Astronomy, K\"onigstuhl 17, 69117, Heidelberg, Germany}
\def\auckland{Dept. of Statistics, The University of Auckland, Private Bag 92019, Auckland 1142, New Zealand}
\def\dark{Dark Cosmology Centre, Niels Bohr Institute, University of
Copenhagen, Juliane Maries Vej 30, 2100 Copenhagen {\O}, Denmark}
\def\nbia{Niels Bohr International Academy, Niels Bohr Institute,
University of Copenhagen, Blegdamsvej 17, 2100 Copenhagen {\O},
Denmark}

\def\breweremail{\tt bj.brewer@auckland.ac.nz}

\def\packard{Packard Research Fellow}

\title[Hierarchical Inference of IMFs]
{The SWELLS Survey. VI. hierarchical inference of the initial mass functions
of bulges and discs}
    
\author[Brewer \etal]{%
  Brendon~J.~Brewer$^{1}$\thanks{\breweremail},
  Philip~J.~Marshall$^{2, 3}$,
  Matthew~W.~Auger$^{4}$,
\newauthor{%
  Tommaso~Treu$^{5}$\thanks{\packard}, 
  Aaron~A.~Dutton$^{6}$,
  Matteo~Barnab\`e$^{7, 8}$
  }
  \medskip\\
  $^1$\auckland\\
  $^2$\kipac\\
  $^3$\oxford\\
  $^4$\cambridge\\
  $^5$\ucsb\\
  $^6$\mpia\\
  $^7$\dark\\ 
  $^8$\nbia
}

\begin{document}
             
\date{}
             
\pagerange{\pageref{firstpage}--\pageref{lastpage}}\pubyear{2012}

\maketitle

\label{firstpage}


\begin{abstract}
The long-standing assumption that the stellar initial mass function
(IMF) is universal has recently been challenged by a number of
observations. Several studies have shown that a ``heavy'' IMF (e.g.,
with a Salpeter-like abundance of low mass stars and thus
normalisation) is preferred for massive early-type galaxies, while
this IMF is inconsistent with the properties of less massive,
later-type galaxies. These discoveries motivate the hypothesis
that the IMF may vary (possibly very slightly) across galaxies and across components of individual
galaxies (e.g. bulges vs discs). In this paper we use a sample of 19
late-type strong gravitational lenses from the SWELLS survey to investigate the
IMFs of the bulges and discs in late-type galaxies. We perform a joint analysis
of the galaxies' total masses (constrained
by strong gravitational lensing) and stellar masses (constrained by
optical and near-infrared colors in the context of a stellar population
synthesis [SPS] model,  up to an IMF normalisation parameter).
Using minimal assumptions apart from the physical constraint that the total stellar mass $m^{*}$
within any aperture must be less than the total mass $\mtot$ within the aperture, we find that the bulges
of the galaxies cannot have IMFs heavier (i.e. implying high mass per unit
luminosity)
than Salpeter, while the disc IMFs are
not well constrained by this data set.
We also discuss the necessity
for hierarchical modelling when combining
incomplete information about multiple astronomical objects.
This modelling approach allows us to place upper limits on
the size of any departures from universality.
More data, including spatially
resolved
kinematics (as in paper V) and stellar population diagnostics over a range of bulge
and disc masses, are needed to robustly quantify how
the IMF varies within galaxies.
\end{abstract}



\section{Introduction}
The stellar initial mass function is a quantity of great interest for
a number of areas in astrophysics, ranging from the understanding of
the microphysics of star formation to the demographics of stars and
galaxies in the Universe.  Owing to its fundamental importance, it is
not surprising that substantial effort is being devoted to measuring
it from astronomical observations.  In the Milky Way the IMF can be
inferred from resolved stellar populations although selection effects
and dynamical and stellar evolution modelling are significant sources
of uncertainty \citep[see][for a recent review]{2010ARA&A..48..339B}. Outside
our Galaxy one has to rely on less direct probes, such as spectral
features in integrated stellar populations \citep[e.g.][]
{2010Natur.468..940V, 2012ApJ...747...69C, 2012ApJ...753L..32S,
2013arXiv1305.2273L}, gravitational lensing \citep[e.g.][]{2010ApJ...709.1195T,
2010ApJ...721L.163A}, and stellar kinematics \citep[e.g.][]{2012Natur.484..485C,
2011arXiv1111.2905D}.

An important indirect way to characterise the IMF in distant galaxies
consists of measuring the total gravitational mass (e.g., from
dynamics, gravitational lensing, X-ray temperature profiles, or some
combination of these) and relating this total mass to the stellar mass
by using some prescription to account for the non-baryonic dark matter
mass. By comparing this gravitationally-measured stellar mass with the
stellar mass obtained from SPS models
constrained by the observed colours of the galaxies, one can infer the
normalisation of the IMF \citep[e.g.][]{2001ApJ...550..212B}. In the past few
years, these methods have provided evidence against a universal IMF.
Early-type galaxies much more massive than the Milky Way seem to
prefer ``heavy'' IMFs, where we use this term to refer to IMFs that
yield mass to light ratios comparable to or larger than those obtained
for a Salpeter IMF \citep[e.g.][]{2013arXiv1306.2635B, 2013ApJ...765....8T, 2012Natur.484..485C,
2012ApJ...752..163S, 2010ApJ...724..511A, 2010ApJ...709.1195T}.
Remarkably, these results are in qualitative agreement with
each other and with trends observed from spectral features. The
normalisation of the IMF appears to be dependent on galaxy mass
or morphology. Analyses of lower-mass or late-type galaxies are
inconsistent with such heavy IMFs \citep[e.g.][]{2001ApJ...550..212B,
2006MNRAS.366.1126C, 2011ApJ...739L..47B}.

\section{The SWELLS Survey}
The Sloan WFC Edge-on
Late-type Lens Survey (SWELLS) survey \citep{2011MNRAS.417.1601T} is a project
aimed at discovering edge-on spiral galaxies acting as gravitational lenses in
the SDSS.
High quality imaging obtained from the HST and Keck telescopes, combined with
dynamical information, provides the opportunity to study the structure of the dark
matter haloes of spiral galaxies, along with the properties of the stellar components
of the lenses.

In a recent paper (Brewer et al. 2012,  hereafter SWELLS-III) we
put forward a simple method for placing upper limits on the IMF normalisation
$\alpha$ in a sample of gravitational lens galaxies from the
SLACS \citep{2006ApJ...638..703B} and SWELLS \citep{2011MNRAS.417.1601T} surveys. The gravitational lensing-derived total
mass (within any aperture) could be measured and compared to the
SPS-derived stellar mass, assuming a Salpeter IMF \citep{1955ApJ...121..161S}. The IMF normalisation
parameter $\alpha$ can be varied but could never be so high that it would
imply a stellar mass that is greater than the measured total mass.
Using this argument, we found that, for the 14 lighter galaxies in the sample
(those with
lens velocity dispersion $\sigma<230$ \kms), a heavy IMF was directly
ruled out by the data, regardless of any assumptions on the non-baryonic
dark matter content; when combined with the previous measurements of a
heavy IMF in more massive galaxies, this result is inconsistent
with a universal IMF.  A possible interpretation of this finding is that
the IMF varies systematically with  galaxy stellar mass. In turn, the
present-day stellar mass of galaxies is known to correlate with the age
and metallicity of their stellar populations, perhaps suggesting that
variations in the IMF reflect differences in the conditions at the time
when the stars were formed.

In this paper, we extend the investigation in SWELLS-III and remedy a number of
shortcomings in the analysis.
The work presented here is complementary to that of
\citet{2013MNRAS.428.3183D}, who re-analyzed a sub-sample of five bulge-dominated
SWELLS systems order to investigate the same question; our approach is
to re-visit the larger sample using an ensemble analysis of the simple,
robust measurements of SWELLS-III.
Specifically, we consider the hypothesis that
the IMF normalisation may vary (possibly very slightly) from galaxy to galaxy, 
and that it may also be different in the bulges and disks of late-type galaxies.
We perform a joint analysis of the
photometric  stellar mass and gravitational total mass measurements of
the SWELLS galaxies in order to infer the IMF normalisation parameter(s)
$\alpha$ of the galaxies' stellar population(s). We make a working
assumption that the IMF  normalisation parameter $\alpha$ may be
different (although possibly close) in  bulges and discs and also across
different objects.
%
Physically, this investigation is motivated by the fact that the stellar
populations of bulges and discs are known to be significantly different
in their star formation histories and chemical abundance patterns,
suggesting that their stars were formed under different conditions.
The presence or absence of differences between the bulge and disc IMFs
could thus offer some clues to the physical
origins of the IMF.  
We note that the analysis presented in this paper follows a general structure
that is common in astrophysics. In a sample of $N$ objects, it may be possible
to measure some property $\{x_i\}$ of each object. However, the measurements
are noisy and so we do not know the property for each object precisely. Despite
this difficulty it is still possible to infer things about the {\it distribution}
of properties present in the population (or at least, in the population as
modulated by the selection function).

The kind of problem is well described
by a Bayesian hierarchical model \citep{2012arXiv1208.3036L}.
These models are becoming
increasingly recognised as a powerful tool in astrophysics, with applications
ranging from exoplanets \citep{extreme_deconvolution}, galaxy evolution \citep{2012AJ....143...90S},
trans-neptunian objects
\citep{loredo}, and fitting of straight lines \citep{kelly}.

This paper is organized as follows. In Section~\ref{sec:notation}, we describe
the notation we will use for the various quantities involved in this study.
In \Sref{sec:data} we define the sample of
galaxies used in this work and the data that will be used for our inferences.
In \Sref{sec:model-onegalaxy} we describe the model
we use to interpret the mass measurements of a single galaxy, and present some
inferences from one system in \Sref{sec:results-onegalaxy}. Then, we 
describe
our hierarchical model for combining the information from all galaxies in the
sample in \Sref{sec:model-hb}, 
and present our final results in \Sref{sec:results-hb}. In \Sref{sec:discuss}
we briefly discuss our results and conclude.


\section{Notation}\label{sec:notation}
Throughout this paper, all masses given are in units of solar masses
$M_{\odot}$. We denote masses by lower-case $m$'s, log-masses (base
10) by upper-case $M$'s, and observed quantities 
by an over-tilde. For
example, an object might have a mass $m$ (in solar masses) and its logarithm is
$\log_{10}(m) = M$. If $M$ is measured it produces a noisy measurement,
\begin{equation}
\widetilde{M} = M + \epsilon,
\end{equation}
where $\epsilon$ is the specific amount of noise in this measurement. We will also have quantities
that are logarithms (to base 10) of masses that have been rescaled by an
unknown factor $\alpha$. These quantities are denoted by curly
upper-case $\mathcal{M}$'s, and correspond to the estimates
of stellar mass made by SPS models:
\begin{equation}
\mathcal{M} = \log_{10}\left(\frac{m}{\alpha}\right) = M - \log_{10}\alpha.
\label{eq:curly_M}
\end{equation}
If $\mathcal{M}$ is estimated under the assumption of a Salpeter
IMF, $\alpha = 1$. Lighter IMFs give lower predicted stellar masses:
a Chabrier IMF \citep{2003PASP..115..763C} corresponds to $\alpha \approx 0.6$. Throughout this paper,
we adopt $\alpha = 0.6$ as the {\it definition} of a Chabrier IMF.
The quantity $\mathcal{M}$ can be considered as the log of the stellar mass
that would be inferred by applying an SPS model that assumes the Salpeter
IMF to be true. For example, suppose a disc has a true mass $m=10^{10} M_{\odot}$
and the appropriate IMF is given by $\alpha=0.5$. Then the application of an
SPS model that assumes a Salpeter IMF would overestimate the mass, giving a
result $\mathcal{M} = \log_{10}\left(\frac{m}{\alpha}\right) = 10.301$.

Our galaxy model is composed of three components -- a
bulge of stars, a disc of stars, and a halo of dark matter -- and the masses
of these three components will be denoted by subscripts b, d, and
h respectively.
We will also consider total mass, denoted with a subscript ${\rm tot}$,
\begin{equation}
\mtot = \mbulge + \mdisc + \mhalo.
\end{equation}
Corresponding logarithmic versions and observed noisy versions will be denoted
by the appropriate upper-case $M$ and an over-tilde as described above.
The fraction of the mass in stars is given by
\begin{equation}
f = \frac{\mbulge + \mdisc}{\mtot}.
\end{equation}
Our modelling ensures that this fraction $f$ lies between 0 and 1. The log of this fraction will be denoted
by the upper case $F$:
\begin{eqnarray}
F &=& \log_{10}(f),
\end{eqnarray}
which lies between $-\infty$ and 0.
Finally, we also define the {\it logit}-transformed version of the stellar mass
fraction $f$, defined by:
\begin{eqnarray}
\ell &=& \ln\left(\frac{f}{1-f}\right).
\end{eqnarray}
The logit transform is useful because $\ell$ can take any real value, while $f$ is constrained to only have values between 0 and 1. This property will be exploited in the full
hierarchical model (Section~\ref{sec:model-hb}).
Note that the stellar mass fractions are defined within the critical curve,
which is an observer-dependent quantity, and not an intrinsic property of the
lens galaxy. Since the mass fractions are nuisance parameters in the context
of studying the IMF, this is not expected have a major impact on our results.

We note that all of the quantities described here are defined for a single
galaxy. As there are 19 lens galaxies in our sample (Section~\ref{sec:data})
there are 19 values for $f$, $m_{\rm tot}$, $\mathcal{M}_b$, etc. Individual
object quantities are denoted by a superscript, for example $F^j$ would be the
logarithm (base 10) of the fraction of the mass that is in stars, for galaxy
$j$.


\section{Data}
\label{sec:data}
For this study we use a subset of 19 systems from the full SWELLS sample
(SWELLS-I, SWELLS-III) that have robust lens models previously published
in SWELLS-III. Specifically, for these 19 lens galaxies we have Singular
Isothermal Ellipsoid models with external shear (SIE+$\gamma$ models).
We also have estimates of the stellar masses for both bulge and disc
components. We integrated the surface brightness models and gravitational lens
models within the
elliptical critical curve to obtain estimates of three quantities:
the bulge mass, the disc mass, and the total mass contained within the critical
curve.
These aperture masses are presented in \Tref{tab:data} and
constitute the data that will be analysed in this paper. Note that all masses
mentioned in this paper refer to the {\it aperture mass integrated
within the critical curve of the lens model}.

The bulge and disc mass measurements in Table~\ref{tab:data} were derived
under the assumption of a Salpeter IMF. Hence, they can be considered as
measurements of the (logarithm of the) stellar masses divided by the
IMF normalisation parameter $\alpha$, as in Equation~\ref{eq:curly_M}.

\begin{table*}
\begin{center}
\begin{tabular}{|l|c|ccc|}
\hline
System & 
$\widetilde{\sigmasie}$ &
$\obsSPSMbulge \pm \errSPSMbulge$ &
$\obsSPSMdisc  \pm \errSPSMdisc$ &
$\obsMtot      \pm \errMtot$ \\
     &   (km/s)  & & &\\
\hline
J0820$+$4847 & 192.3 & 10.65 $\pm$ 0.09 & 10.15 $\pm$ 0.08 & 10.71 $\pm$ 0.047 \\
J0822$+$1828 & 185.7 & 10.54 $\pm$ 0.16 &  9.85 $\pm$ 0.16 & 10.64 $\pm$ 0.018 \\
J0841$+$3824 & 251.7 & 11.07 $\pm$ 0.09 & 10.01 $\pm$ 0.09 & 11.15 $\pm$ 0.015 \\
J0915$+$4211 & 196.0 & 10.38 $\pm$ 0.11 & 10.50 $\pm$ 0.13 & 10.61 $\pm$ 0.009 \\
J0930$+$2855 & 344.0 & 11.23 $\pm$ 0.17 & 10.35 $\pm$ 0.17 & 11.61 $\pm$ 0.006 \\
J0955$+$0101 & 238.3 & 10.60 $\pm$ 0.09 & 10.21 $\pm$ 0.09 & 10.93 $\pm$ 0.023 \\
J1021$+$2028 & 162.8 & 10.61 $\pm$ 0.18 &  9.48 $\pm$ 0.12 & 10.30 $\pm$ 0.046 \\
J1029$+$0420 & 212.6 & 10.85 $\pm$ 0.10 &  9.72 $\pm$ 0.09 & 10.82 $\pm$ 0.012 \\
J1032$+$5322 & 251.8 & 10.80 $\pm$ 0.08 & 10.38 $\pm$ 0.10 & 11.06 $\pm$ 0.009 \\
J1103$+$5322 & 224.1 & 10.88 $\pm$ 0.09 & 10.55 $\pm$ 0.10 & 11.03 $\pm$ 0.011 \\
J1111$+$2234 & 229.4 & 11.16 $\pm$ 0.14 & 10.69 $\pm$ 0.15 & 11.18 $\pm$ 0.005 \\
J1117$+$4704 & 218.4 & 10.75 $\pm$ 0.16 & 10.37 $\pm$ 0.12 & 10.88 $\pm$ 0.008 \\
J1135$+$3720 & 207.4 & 10.21 $\pm$ 0.13 & 10.71 $\pm$ 0.14 & 10.79 $\pm$ 0.042 \\
J1203$+$2535 & 189.3 & 10.56 $\pm$ 0.08 & 10.05 $\pm$ 0.09 & 10.63 $\pm$ 0.028 \\
J1251$-$0208 & 205.0 & 10.76 $\pm$ 0.07 & 10.26 $\pm$ 0.08 & 10.95 $\pm$ 0.012 \\
J1313$+$0506 & 174.7 & 10.67 $\pm$ 0.10 &  9.11 $\pm$ 0.11 & 10.44 $\pm$ 0.124 \\
J1331$+$3638 & 248.9 & 10.76 $\pm$ 0.10 &  9.54 $\pm$ 0.08 & 10.95 $\pm$ 0.018 \\
J1703$+$2451 & 209.6 & 10.21 $\pm$ 0.06 &  9.92 $\pm$ 0.09 & 10.47 $\pm$ 0.010 \\
J2141$-$0001 & 190.3 & 10.65 $\pm$ 0.10 & 10.37 $\pm$ 0.08 & 10.77 $\pm$ 0.055\\
\hline
\end{tabular}
\end{center}
\caption{Properties of the 19 galaxies in the subsample of SWELLS
discussed in this paper. The quantities given are the measured value of
the lensing velocity dispersion $\sigmasie$ along with the bulge,
disc and total masses within the critical curve. These values are the ``data''
that we will analyse in this paper.
\label{tab:data}}
\end{table*}

\begin{figure*}
\begin{center}
\includegraphics[width=0.9\linewidth]{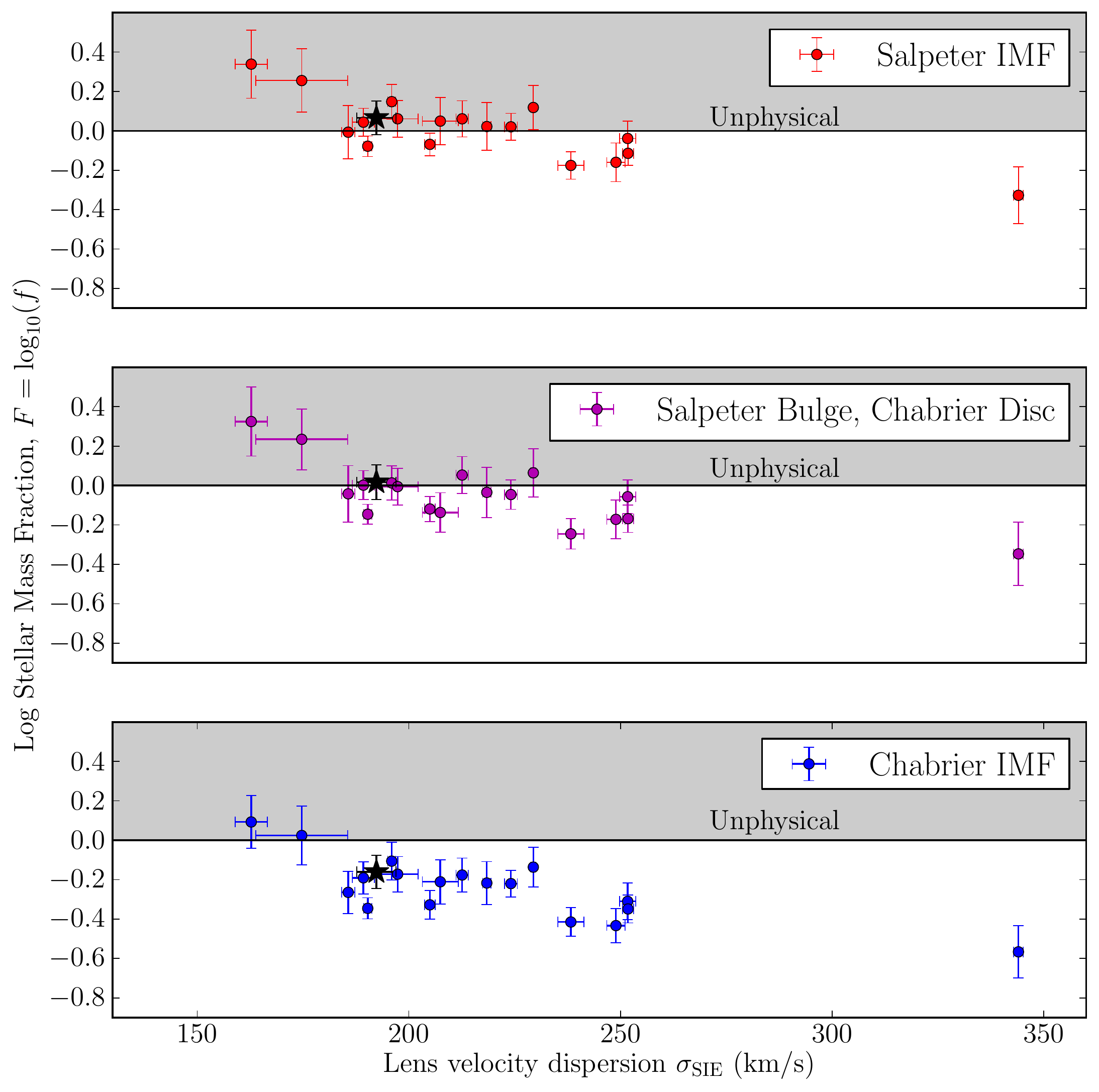}
\caption{A comparison of the ``observed'' stellar mass fraction 
as a function of lens velocity dispersion for three different IMF
scenarios.  The vertical  error bars are approximate, and were computed
by calculating the uncertainty in the ratio of the stellar mass to the
total mass {\it assuming those two quantities are independent}. The top
two panels are similar to plots from SWELLS-III, while the middle panel
is new to this paper. {\bf Top}: Salpeter IMF, {\bf Middle}: Salpeter
IMF for bulge, Chabrier IMF for disc, {\bf Bottom}: Chabrier IMF. The galaxy
denoted with the black star symbol is J0820$+$4847, which is the object
studied in Section~\ref{sec:model-onegalaxy}.}
\label{fig:trend1}
\end{center}
\end{figure*}

In Figure~\ref{fig:trend1}, we plot, as a function of the estimated
value of the gravitational lens
velocity
dispersion parameter $\widetilde{\sigmasie}$, the ``observed'' mass
fraction in stars,
calculated na\"{\i}vely by dividing the measured stellar mass by the measured
total mass:
\begin{eqnarray}
F \approx \log_{10}\left(10^{\widetilde{\SPSMbulge} +
\widetilde{\SPSMdisc}}\right) 
- \widetilde{\Mtot}
\end{eqnarray}
In the top panel, we see the assumption of a Salpeter IMF leads to a 
prediction of high stellar mass fraction, and many of the galaxies in
our sample appear in the unphysical region $F > 0$ (or $f > 1$). Thus, this
data set seems to provide some evidence against the Salpeter IMF, and
``heavier'' IMFs,
for this sample. Our
goal in the remainder of this paper is to formalise this intuition by
carefully performing the inference and obtaining constraints
on hypotheses about the IMF normalisations $\alpha$ for the bulges and discs
in this sample. This will be achieved by a Bayesian
analysis of the data given in Table~\ref{tab:data}. This analysis should
be considered as superseding the one presented in SWELLS-III.


\section{A Single Galaxy}\label{sec:model-onegalaxy}


\subsection{Noise-Free Argument}
In this section we describe a simple argument for constraining the IMF
normalisation parameters $\alphabulge$ and $\alphadisc$ in a single
galaxy based on an SPS mass measurement (that assumed a Salpeter IMF)
and a total mass measurement. This simple argument provides the fundamental
reason for why this data set can constrain the IMF, free of the later
probabilistic complications.

Suppose that the log of the total mass of a galaxy (within some
aperture, in our case the elliptical critical curve of each system's lens model) has been measured
exactly (without
noise), giving the result
\begin{eqnarray}
\Mtot &=& \log_{10}\left(\mbulge + \mdisc + \mhalo\right).
\end{eqnarray}
Suppose also that the bulge mass and the disc mass 
{\it within the same aperture} have also been measured exactly from the
galaxy photometry and the application of an SPS model that assumes a
Salpeter IMF. This is equivalent to having obtained the values of the
following ``SPS masses:''
\begin{eqnarray}
\SPSMbulge &=& \log_{10}\left(\frac{\mbulge}{\alphabulge}\right) \\
\SPSMdisc &=& \log_{10}\left(\frac{\mdisc}{\alphadisc}\right).
\end{eqnarray}
In terms of these quantities, the total stellar mass (bulge plus
disc) within the aperture is:
\begin{eqnarray}
\mbulge + \mdisc &=& \alphabulge 10^{\SPSMbulge} + \alphadisc 10^{\SPSMdisc}.
\end{eqnarray}
The key argument is that the total stellar mass cannot be greater
than the total mass including dark matter. This leads to the following
constraint:
\begin{eqnarray}
\mbulge + \mdisc &\leq& \mtot\\
\alphabulge 10^{\SPSMbulge} + \alphadisc 10^{\SPSMdisc} &\leq& 10^{\Mtot}.
\label{eq:constraint}
\end{eqnarray}
Equation~\ref{eq:constraint} can be interpreted as a linear constraint
on $\alphabulge$ and $\alphadisc$. This constraint, produced under the (false)
assumption of noise-free measurements, is plotted as the dotted green line in
Figure~\ref{fig:simple:alphas}, alongside more justified conclusions
derived henceforth.


\subsection{Inference in the Presence of Noise}
\label{sec:noerror}
The assumption of noise-free measurements above is of course invalid. We
now relax this assumption in order to obtain a probabilistic version of
the constraint Equation~\ref{eq:constraint}.

Bayesian inference provides the framework for inferring the values of
unknown model parameters. This is done by assuming a prior distribution
$\pr(\theta)$ over the parameter space (where $\theta$ denotes the unknown
parameters) and updating this to a posterior distribution using Bayes' rule:
\begin{eqnarray}
\pr(\theta|D) \propto \pr(\theta)\pr(D|\theta).
\end{eqnarray}

The first ingredient for the analysis is a choice of prior distribution
$\pr(\theta)$ describing initial uncertainty about the parameter values.
The second ingredient for the analysis is a choice of ``sampling distribution''
$\pr(D|\theta)$ which describes our assumptions about how the data came about,
or our predictions about data we would be likely to observe if we happened to
know the values of the unknown parameters. When the data are fixed the sampling
distribution is a function of the parameters only, called the likelihood
function. If the set of possible values for $\theta$ and $D$ is continuous then $\pr(\theta)$ and
$\pr(D|\theta)$ are probability density functions (PDFs).

At this point we note
a common misconception: that priors are subjective and sampling distributions
are somehow measurable. In fact, both priors and sampling distributions
describe prior beliefs (which may be informed by previous data).
The prior $\pr(\theta)$ models an agent's
prior beliefs about the values of the parameters, and the sampling distribution
$\pr(D|\theta)$ models the agent's prior beliefs about how the data are
related to the parameters. Without prior knowledge of a relationship, inference
is not possible.

The final ingredients in the analysis are the actual data $D$ and numerical
methods (such as Markov Chain Monte Carlo [MCMC]) for computing the posterior
distribution $\pr(\theta|D)$ or summaries thereof.

\subsubsection{Data}
In our case, the data are the values of the three measured
quantities:
\begin{eqnarray}
D = \{\obsSPSMbulge, \obsSPSMdisc, \obsMtot\}.
\end{eqnarray}
In principle, a more ``raw'' version of the data, such as pixel values, should
be used. However, this is prohibitive in many applications.

For concreteness we will
use the galaxy \onegalaxy to demonstrate the technique. This galaxy
has the following mass measurements within its critical curve:
\begin{equation}
\begin{array}{lllll}
\obsSPSMbulge &=& 10.65\\
\obsSPSMdisc &=& 10.15\\
\obsMtot &=& 10.71.
\end{array}
\end{equation}
The uncertainties on these measurements are:
\begin{equation}
\begin{array}{lllll}
\errSPSMbulge &=& 0.09\\
\errSPSMdisc &=& 0.08\\
\errMtot &=& 0.05.
\end{array}
\end{equation}
These estimates and uncertainties are the result of SPS modelling (in
the case of $\SPSMbulge$ and $\SPSMdisc$) and gravitational lens
modelling (in the case of $\Mtot$).

\subsubsection{Sampling Distribution}
We make the usual Gaussian
(normal) assumption for the measured quantities given the true
quantities:
\begin{equation}
\def\arraystretch{-1}%
\begin{array}{lllll}
\obsSPSMbulge &\sim& \textnormal{Normal}\left(\SPSMbulge, \errSPSMbulge^2\right)\\
\obsSPSMdisc  &\sim& \textnormal{Normal}\left(\SPSMdisc, \errSPSMdisc^2\right)\\
\obsMtot      &\sim& \textnormal{Normal}\left(\Mtot, \errMtot^2\right).
\end{array}
\label{eq:likelihood}
\end{equation}

\subsubsection{Parameters and Priors}
The unknown parameters of interest to be inferred are $\alphabulge$ and
$\alphadisc$. However, the sampling distributions in \Eref{eq:likelihood} depend
on the values of the unknown {\it nuisance} parameters $\SPSMbulge$, $\SPSMdisc$
and $\Mtot$, so they must also be included as model parameters. For
convenience, rather than using $\Mtot$ as a parameter we will parameterise
in terms of $\logfstar$, which so our five unknown parameters are:
\begin{eqnarray}
\theta = \left\{\alphabulge, \alphadisc,
\SPSMbulge, \SPSMdisc, \logfstar\right\}
\end{eqnarray}
Therefore, the final line of the sampling distributions (\Eref{eq:likelihood})
will need to be written in terms of these parameters.
Note that:
\begin{eqnarray}
\mtot &=& \frac{\mbulge + \mdisc}{f}
\end{eqnarray}
Which implies:
\begin{eqnarray}
\Mtot &=& \log_{10}\left(\alphabulge 10^{\SPSMbulge} +
			\alphadisc 10^{\SPSMdisc}\right) - \logfstar
\end{eqnarray}
Therefore, the final line of the likelihood (\Eref{eq:likelihood}) can be
rewritten in terms of the parameters $\theta$:
\begin{eqnarray}
\obsMtot \sim \textnormal{Normal}
\left(\log_{10}\left(\alphabulge 10^{\SPSMbulge} +
		\alphadisc 10^{\SPSMdisc}\right) - \logfstar, \errMtot^2\right)
\end{eqnarray}

To assign priors, we can use broad distributions that allow a
wide range of plausible values. For $\alphabulge$ and $\alphadisc$ we use a
plausible range from 0.3 to 2. Let us also (for now) assume that we know ``almost
nothing'' about the masses of galaxies, but guess that each component's
mass (assuming a Salpeter IMF) lies somewhere between $10^5$ and
$10^{15}$ solar masses. This is a very broad prior range and a much narrower
prior would be astrophysically justified, however, the prior range does not
have much of an effect because these quantities are directly measurable and the
posterior distribution will be much narrower than the prior.

This prior information is encoded by the
following choices for the prior distribution (for four of the five
parameters):
\begin{eqnarray}
\alphabulge &\sim& \textnormal{Uniform}(0.3, 2)\\
\alphadisc  &\sim& \textnormal{Uniform}(0.3, 2)\\
\SPSMbulge  &\sim& \textnormal{Uniform}(5, 15)\\
\SPSMdisc   &\sim& \textnormal{Uniform}(5, 15)\label{eq:prior1}
\end{eqnarray}
The final parameter that requires a prior distribution is $\logfstar$.
This variable naturally has an upper limit of zero, and we shall assign
a uniform prior:
\begin{eqnarray}
\logfstar   &\sim& \textnormal{Uniform}(-5, 0).
\end{eqnarray}
Other quantities that may be of
interest are considered derived quantities and can be obtained from the
formulae given in Section~\ref{sec:notation}.

The choice of uniform priors, particularly for $\alphabulge$ and $\alphadisc$,
is not intended to be a realistic description of the best judgment of an IMF
expert. Rather, the uniform prior is a convenient choice that allows us to
easily see the effect of the current data set in the results.

\subsubsection{Computation}
To obtain the marginal
distributions for $\alphabulge$ and $\alphadisc$ we implemented this
model in the \jags\footnote{Just Another Gibbs 
Sampler:\newline \tt http://mcmc-jags.sourceforge.net/}~program for 
MCMC \citep{jags}. \jags~was chosen
because it is straightforward to implement complex Bayesian models using
its model specification language. For this study, this attribute was
more important than computational efficiency.


\section{Results: the bulge and disc IMFs in a single galaxy}
\label{sec:results-onegalaxy}

\begin{figure}
\begin{center}
\includegraphics[width=0.95\linewidth]{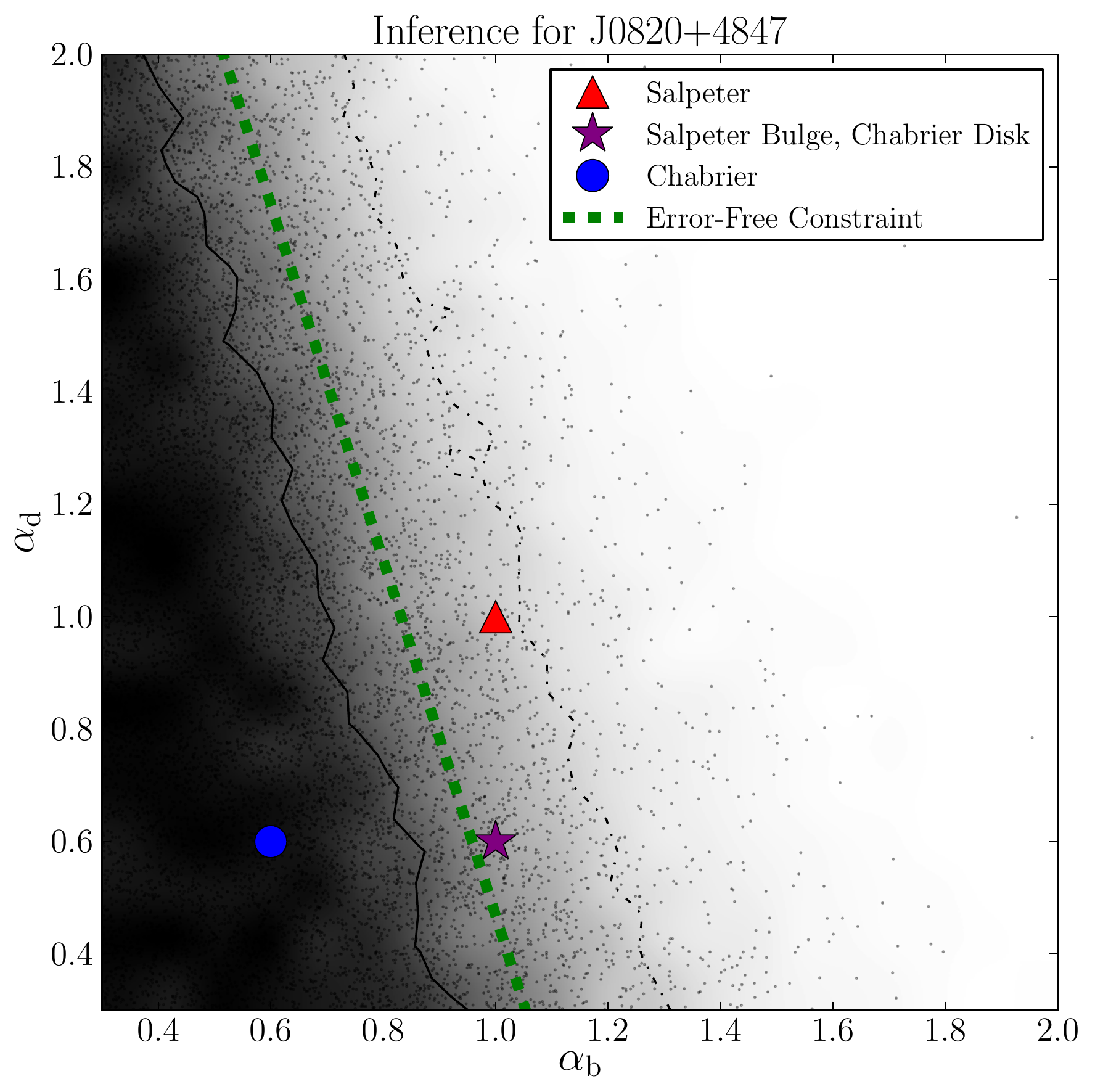}
\end{center}
\caption{The posterior PDF for the bulge and disc IMF normalisation parameters
$\alphabulge$ and $\alphadisc$ for \onegalaxy. Darker regions have higher probability
density.
If noise-free measurements
were available, the green dotted line would provide a hard limit. The presence
of noise has smoothed out the constraint, so that more massive
IMFs become less plausible in a gradual manner. Note that the bulge IMF is
more constrained than the disc IMF. This occurs because the bulge has more
mass within the critical curve than the disc.
The inner contour encloses 68\% of the probability and
the outer contour encloses 95\%. The hypothesis that both bulge and disc IMF
normalisations are Salpeter is indicated by the red triangle
at (1, 1). Similarly, the purple star at (1, 0.6) and the blue circle
at (0.6, 0.6) denote other
hypotheses about the IMF of the bulge and the disc in \onegalaxy.
\label{fig:simple:alphas}}
\end{figure}

\Fref{fig:simple:alphas} shows the posterior PDF
for the bulge and disc IMF normalisations, $\alphabulge$ and
$\alphadisc$, given the \onegalaxy data. For clarity, the dotted line
shows the constraint that would be obtained from
\Eref{eq:constraint} if the measurements had zero uncertainty
(Section~\ref{sec:noerror}; note the perfect degeneracy between the bulge
and disc IMF normalisations). The posterior is
approximately uniform below the constraint line, just like the prior.
As expected, the probabilistic model has softened the hard constraint. We
see that $\alphabulge$ is more constrained than $\alphadisc$, a result
of the fact that this lens is dominated by the bulge, within its critical
curve (as are most of the SWELLS lenses).
We find that, under these
assumptions, and marginalising over all other parameters, $\alphabulge <
1.10$ with 95\% probability. For comparison, the prior probability of the
hypothesis $\alphabulge < 1.10$ was 47\%.


\section{A hierarchical model}
\label{sec:model-hb}
We now consider the simultaneous analysis of multiple
galaxies in order to produce stronger constraints than those produced
in the single system \onegalaxy and shown in
Figure~\ref{fig:simple:alphas}. It is tempting to simply repeat the
analysis of Section~\ref{sec:model-onegalaxy} for each system and take the
product of the resulting marginal likelihood as a function of $\alphabulge$ and
$\alphadisc$. However, such an approach would produce innaccurate
results due to the following hidden assumptions. Firstly, in this
approach we would be implicitly assuming that there is one $\alphabulge$
value that applies to all bulges and one $\alphadisc$ value that applies
to all discs. Secondly (and more importantly), the model in
Section~\ref{sec:model-onegalaxy}
assumes a broad prior on the logarithm of the stellar mass fraction ($F$).
Applying this model to every galaxy is equivalent to assuming the prior on
the set of $F$-values ($\{F^j\}$) is independent. This assumption is
problematic because it implies that the $F$ values are all very likely to
be scattered across the whole range $[-5, 0]$ and are very unlikely to be
concentrated together.

Thus, the prior probability that (for example) all objects have a high
stellar mass fraction is unreasonably low, and the model will favour
lower values for the IMF normalisation. Inspection of \Fref{fig:trend1}
suggests that a universal Salpeter IMF should be disfavoured; however
the strength of this conclusion will depend strongly on how plausible it
is that {\it most galaxies in the sample have high stellar mass
fractions}. Applying the
\Sref{sec:model-onegalaxy} analysis independently to each object
implies a very small prior probability for {\it
all} galaxies having approximately the same high $\logfstar$ value, while we
have no good reasons to be sceptical of this possibility.
In
other words, such an analysis ignores our belief that some common
physical framework is responsible for the formation and evolution of all
galaxies. The result from applying this strategy is shown in Appendix~\ref{sec:wrong}.

To overcome these problems, we introduce an alternative hierarchical
Bayesian model. We allow each galaxy to have its own value for
$\alphabulge$ and its own value for $\alphadisc$, but rather than
assigning independent Uniform$(0.3, 2)$ priors for these parameters, we
assume that the prior would be a lognormal distribution, if only we knew
the appropriate mean and variance. However, we do not know the appropriate
mean and variance, so we allow those to be unknown parameters as well.
There are several ways to parameterise a lognormal distribution. Throughout
this paper we use the median value $\mu$ of the variable and the standard
deviation of the natural logarithm of the variable, denoted by $\sigma$.
The lognormal density for a variable $x$ is given by:
\begin{eqnarray}
\pr(x|\mu, \sigma) &=& 
\frac{1}{x\sigma\sqrt{2\pi}}
\exp\left[
-\frac{1}{2\sigma^2}
\left(\ln(x) - \ln(\mu)\right)^2
\right]
\end{eqnarray}
for $x \geq 0$.
When applied to the bulge IMF normalisations,
the hierarchical prior increases the prior probability
that the $\alphabulge$ values are very similar to each other across the
population and encodes our belief that the physics of star formation is not
wildly different in different galaxies. The median and variance of this
phenomenological model for the distribution of bulge IMFs are to be
inferred from the data. Similarly, we take the $\alphadisc$ parameters
of each disc to be drawn from a different lognormal distribution.
We now have four
{\it hyperparameters} that represent our simple model for the distribution
of bulge and disk IMFs in the population.
It is to these hyperparameters -- the medians and variances of the conditional
prior PDFs for the $\alphabulge$ and $\alphadisc$ parameters -- we assign vague, 
``uninformative'' priors. 
The $\{\alphabulge^j\}$ and $\{\alphadisc^j\}$ variables are not the only
quantities whose prior should be modified when the model is extended to
multiple galaxies. We shall assume the bulge ``masses'' $\{\SPSMbulge^j\}$ 
were drawn from a
common normal distribution, whose mean $\SPSMbulgemedian$ and width
$\SPSMbulgewidth$ we
hope to infer from the sample. This corresponds to the assumption that
our galaxies' bulges are similar in their mass, which makes sense given
that they were all {\it selected} in the same way
\citep{2011MNRAS.417.1601T}. We make the same assumption for the discs:
that they are similar in {\it their} mass. This introduces additional
hyperparameters $\SPSMdiscmedian$ and $\SPSMdiscwidth$.

Each of our galaxies has an $F$ value, the logarithm (base 10) of the
fraction of the mass that is in stars (within its critical curve). Rather
than applying a simple hierarchical structure to the prior on the $\{F^j\}$
variables, the apparent trend in Figure~\ref{fig:trend1} motivates the
assumption that the $F$ values for the galaxies vary as a function of
$\widetilde{\sigmasie}$. The hard upper limit of $F=0$ complicates matters,
so we work in the logit basis where $\ell = \ln\left[f/(1-f)\right]$. We shall assume
that the mass fractions are scattered around the following relationship
as a function of $\log_{10}\left(\widetilde{\sigmasie^j}\right)$:
\begin{eqnarray}
\ell &=& \ell_0 + \ell_1\log_{10}\left[\frac{\widetilde{\sigmasie^j}}{162.8\textnormal{ km/s}}\right].
\end{eqnarray}
In this relationship, $\ell_0$ is the ``initial'' value of the trend line,
$\ell_1$ is the gradient, and 162.8 km/s is the lowest
$\widetilde{\sigmasie^j}$ value in the data set, where $\ell=\ell_0$ applies.
This assumption introduces hyperparameters $\ell_0$, $\ell_1$ and
$\sigma_\ell$ into the model, where the latter is the degree of scatter around
the assumed trend line. We choose vague priors for $\ell_0$ and $\sigma_\ell$.
For $\ell_1$ (the slope of the trend), we choose a heavy tailed Cauchy (Lorentzian)
distribution centered at 0 and with a scale of 10. This models a prior assumption
that $\ell_1$ is probably small in magnitude (such that we would not expect
$\ell$ to vary by more than 10 over an order-of-magnitude range of
$\widetilde{\sigmasie^j}$), but the
heavy tails allow this assumption to be incorrect if the data warrant it.

\begin{figure}
\centering\includegraphics[width=0.95\linewidth]{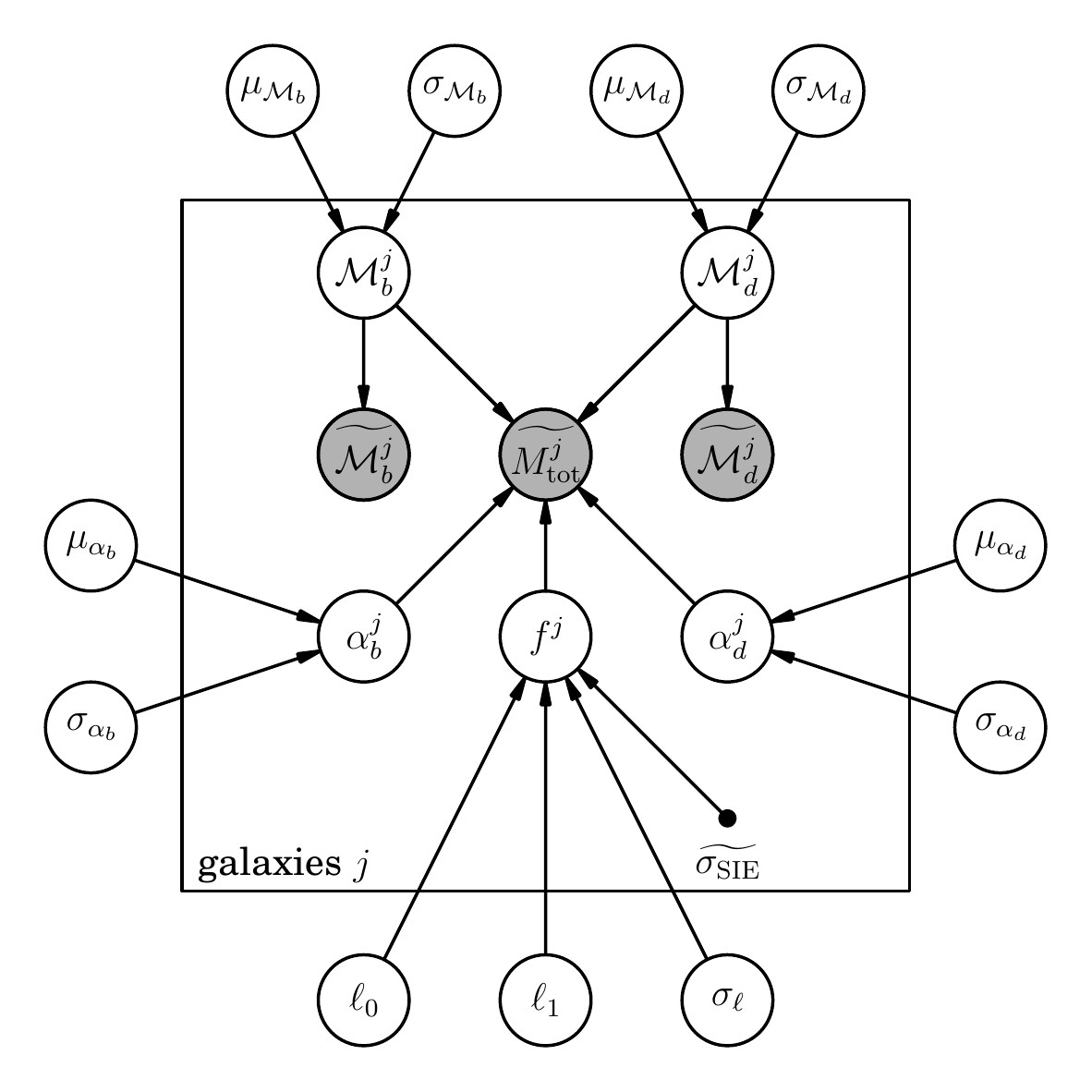}
\caption{Probabilistic graphical model of our hierarchical model, produced using the \daft~package by Dan Foreman-Mackey and
David Hogg, available from
\texttt{http://daft-pgm.org/}.
The equations illustrated by this figure are
given in Table~\ref{tab:everything}. The box can be interpreted as a `for' loop,
indicating quantities that are defined for each lens galaxy in the sample.
Shaded nodes denote observed quantities (aka ``data'')
whose values are conditioned on in the inference. The small black dot indicates
quantities that are fixed and known, but considered as ``prior information''
rather than ``data'' (i.e. no sampling distribution is specified).}
\label{fig:pgm}
\end{figure}

The prior PDF for all hyperparameters and parameters, and the
sampling distribution for the data given the parameters, is given in
\Tref{tab:everything}. The structure of this hierarchical model is depicted
graphically in
\Fref{fig:pgm}. This diagram can be understood as a recipe for generating
simulated data sets by working from the outer nodes (hyperparameters) to the
innermost data nodes. The first step would be to generate hyperparameter
values from their prior, to describe the central tendencies and diversities
of the galaxies' properties. Then, individual galaxy properties would be
drawn from their distributions given the hyperparameters. Finally, data values
would be generated from the sampling distributions given the
individual galaxy properties. A recipe for simulating data (starting by simulating
parameter values from their prior) is equivalent to a fully specified Bayesian
model. In Section~\ref{sec:results-hb}
we present the results from the SWELLS sample.

\setlength{\tabcolsep}{0cm}
\begin{table*}
\begin{tabular}[scale=0.7]{|ccc|}
\hline
{\bf Parameter} & {\bf Meaning} & {\bf Probability Distribution}\\
\hline
{\it Population Hyperparameters}\\
\hline
$\alphabulgemedian$                    & Median of bulge IMFs                                  & Uniform(0.3, 2) \\
$\alphabulgewidth$                     & Diversity of bulge IMFs                               & LogUniform(0.001, 1)  \\
$\alphadiscmedian$                     & Median of disc IMFs                                   & Uniform(0.3, 2) \\
$\alphadiscwidth$                      & Diversity of disc IMFs                                & LogUniform(0.001, 1)  \\
$\SPSMbulgemedian$                     & Median of bulge masses                                & Uniform(5, 15)  \\
$\SPSMbulgewidth$                      & Diversity of bulge masses                             & LogUniform(0.001, 1)  \\
$\SPSMdiscmedian$                      & Median of disc masses                                 & Uniform(5, 15)  \\
$\SPSMdiscwidth$                       & Diversity of disc masses                              & LogUniform(0.001, 1)  \\
$\lfstarintercept$                     & Median of $\lfstar$ (logit-transformed $\fstar$) at $\widetilde{\sigmasie} = 162.8$ km/s & Uniform(-10, 10)\\
$\lfstargradient$                      & Gradient of $\lfstar$ trend with $\log_{10}\left(\widetilde{\sigmasie}\right)$ & Cauchy(0, 10) \\
$\lfstarwidth$                         & Scatter of $\lfstar$ values around trend line            & LogUniform(0.001, 1)  \\
\hline 
{\it Individual Galaxy Parameters}\\
\hline
$\alphabulge^j$                        & Bulge IMF             & LogNormal($\alphabulgemedian$, $\alphabulgewidth^2$)\\
$\alphadisc^j$                         & Disc IMF              & LogNormal($\alphadiscmedian$, $\alphadiscwidth^2$)\\
$\SPSMbulge^j$                         & log$_{10}$(``Salpeter'' Bulge mass)            & $\textnormal{Normal}(\SPSMbulgemedian, \SPSMbulgewidth^2)$\\
$\SPSMdisc^j$                          & log$_{10}$(``Salpeter'' Disc mass)           & $\textnormal{Normal}(\SPSMdiscmedian, \SPSMdiscwidth^2)$\\
$\ln\left(\frac{f^j}{1 - f^j}\right)$ & logit-transformed stellar mass fraction & $\textnormal{Normal}\left(\ell_0 + \ell_1\log_{10}\left[\frac{\widetilde{\sigmasie^j}}{162.8\textnormal{ km/s}}\right], \sigma_\ell^2\right)$\\
\hline 
{\it Data}\\
\hline
$\obsSPSMbulgej$ & SPS bulge mass measurement     & $\textnormal{Normal}(\SPSMbulge^j, (\errSPSMbulge^j)^2)$\\
$\obsSPSMdiscj$  & SPS disc mass measurement      & $\textnormal{Normal}(\SPSMdisc^j, (\errSPSMdisc^j)^2)$\\
$\obsMtotj$      & Lensing total mass measurement & $\textnormal{Normal}\left(\log_{10}\left(\alphabulge^j 10^{\SPSMbulge^j} +
			\alphadisc^j 10^{\SPSMdisc^j}\right) - \logfstar^j, (\errMtot^j)^2\right)$\\
\end{tabular}
\caption{All the parameters in our hierarchical model, with their
assigned prior PDFs or sampling distributions. In our notation for a lognormal
distribution, the first entry gives the {\it median} or central value of the
variable, and the second entry gives the variance (standard deviation squared) of
the natural log of the variable. However, the variance is usually written
as $\sigma^2$ so the standard deviation can be read easily.
The total number of unknown parameters and hyperparameters is 106 and the total
number of measurements is 57 (not including noise variances on the measurements,
which are counted as prior information).
\label{tab:everything}}
\end{table*}

%
%

\section{Results: the disc and bulge IMFs in the SWELLS sample}
\label{sec:results-hb}

\begin{figure}
\begin{center}
\includegraphics[width=0.95\linewidth]{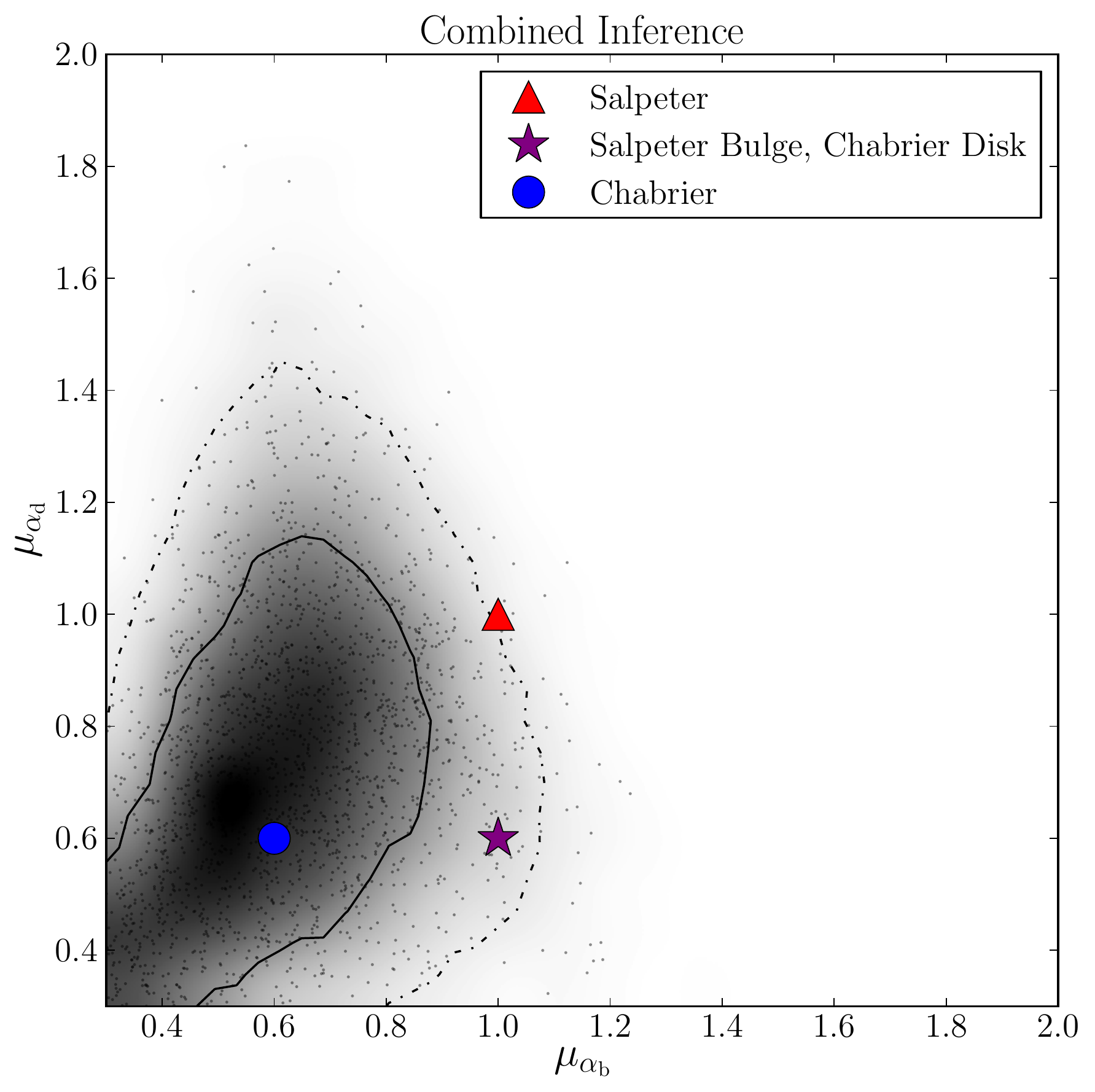}
\caption{Samples from the posterior PDF for the median IMF normalisation
parameters in the  context of our hierarchical model,
$\alphabulgemedian$ and $\alphadiscmedian$, given 19 SWELLS galaxies'
mass measurements. The inner contour encloses 68\% of the probability and
the outer contour encloses 95\%. The hypothesis that both bulge and disc IMF
normalisations are centered around Salpeter is indicated by the red triangle
at (1, 1). Similarly, the purple star at (1, 0.6) and the blue circle
at (0.6, 0.6) denote other
hypotheses about the typical IMFs of bulges and discs.} 
\label{fig:hierarchicalalpha:meanalphas}
\end{center}
\end{figure}

\begin{figure*}
\begin{center}
\includegraphics[width=0.9\linewidth]{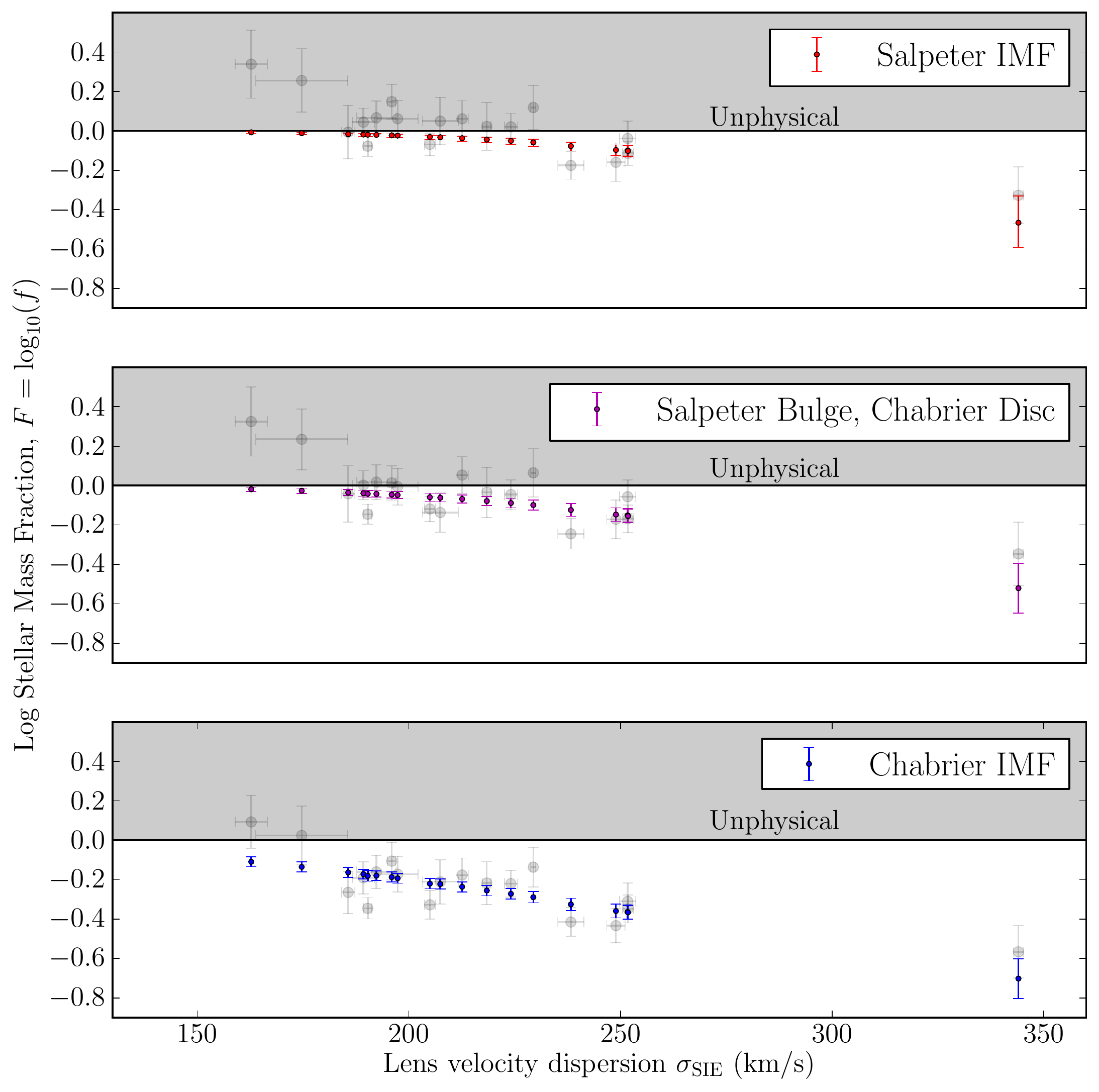}
\caption{The {\it inferred} stellar mass fractions $\logfstar$ from the
inference, for comparison with \Fref{fig:trend1}. The ``observed''
(and often unphysical) $\logfstar$ values are plotted as grey points, while the
hierarchically-inferred $\logfstar$ values are shown in colour.
Top: Salpeter IMF, Middle: Salpeter IMF for bulge, Chabrier IMF for
disc, IMF: Bottom: Chabrier IMF. \label{fig:hierarchical:f}}
\end{center}
\end{figure*}

The hierarchical model specified in Section~\ref{sec:model-hb}
was implemented and run in \jags, producing posterior samples for all of the
hyperparameters and parameters listed in Table~\ref{tab:everything}.
To investigate conclusions similar to those for \onegalaxy in
Figure~\ref{fig:simple:alphas},
we plot the marginal posterior PDF for $\alphabulgemedian$ and
$\alphadiscmedian$ in \Fref{fig:hierarchicalalpha:meanalphas}. We see that in
this sample, the mass measurements tend to support light IMFs in both
the disc and the bulge components. The constraint on the bulge IMF is
marginally tighter than the disc constraint, as we might expect for these
brighter, more massive
galaxy components. Heavy IMFs are somewhat disfavoured for both
components in these galaxies: marginalising, we find that
$\pr(\alphabulgemedian < 1) \approx 96.6\%$ while $\pr(\alphadiscmedian < 1)
\approx 81.6\%$. The prior probability of each of these hypotheses was
41.2\%. In other words, this data strongly disfavours the hypothesis that
the centroid of the distribution of $\alpha$ values is greater than Salpeter (1).
However, the assumption of a near-universal IMF
($\alphabulgemedian \approx \alphadiscmedian$ and both
$\alphabulgewidth$ and $\alphadiscwidth$ are small)
is compatible with the data.

\begin{figure}
\begin{center}
\includegraphics[width=0.9\linewidth]{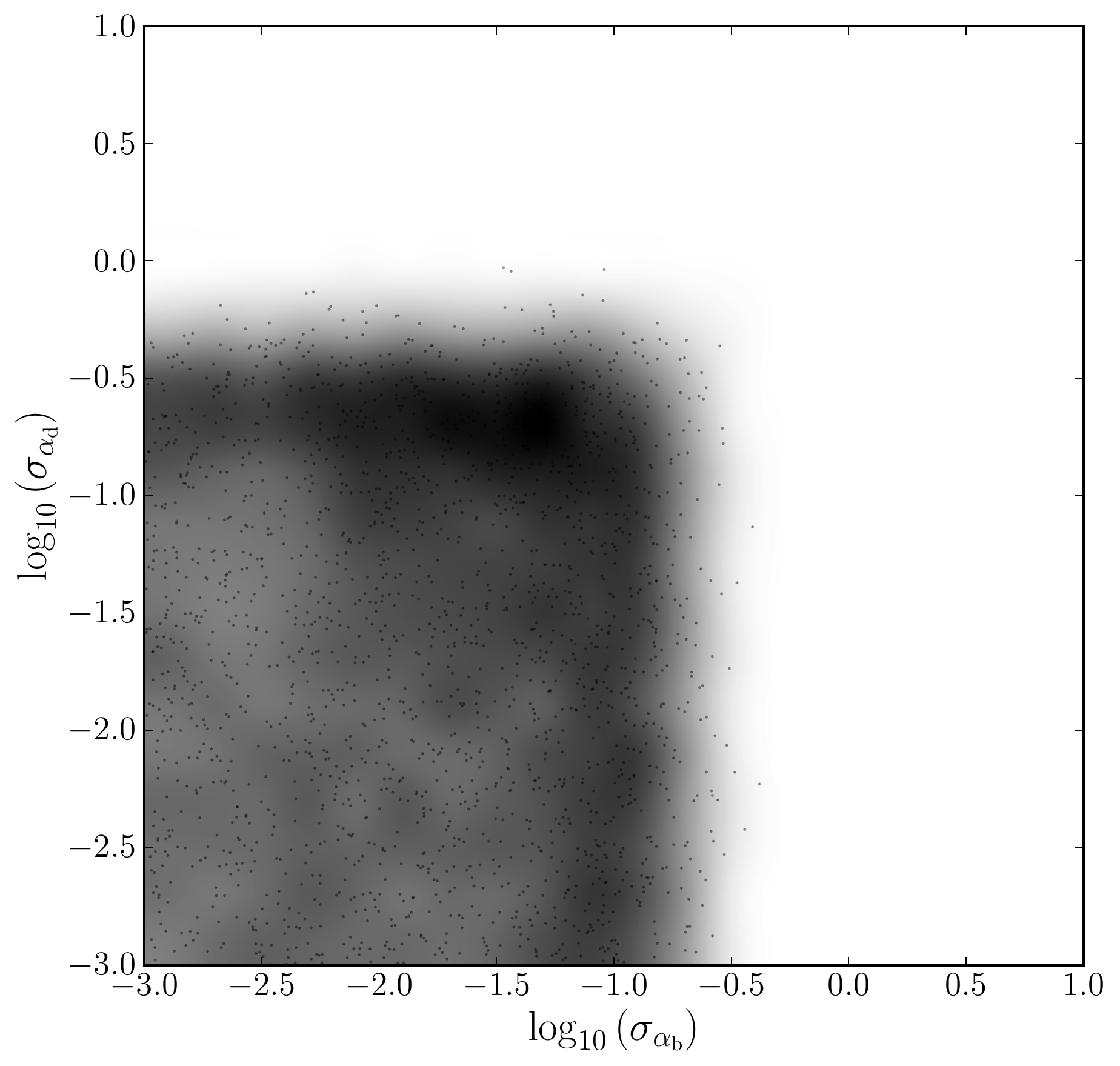}
\caption{Samples from the posterior PDF for the 
width parameters, $\alphabulgewidth$ and  $\alphadiscwidth$, of the
galaxy component IMF normalisation distributions, given 19 SWELLS
galaxies' mass measurements. Small values of $\alphabulgewidth$ imply
a situation where the IMFs of bulges are close to universal, and this
is not ruled out by the data. This is also the case with
$\alphadiscwidth$. The prior was uniform over the square region plotted here,
so the data merely rule out high values of the width parameters.} 
\label{fig:hierarchical:widthalphas}
\end{center}
\end{figure}

Figure~\ref{fig:hierarchicalalpha:meanalphas} shows the posterior distribution for the
median IMF normalisations for bulges and disks, but we have assumed that
there is some (possibly small) diversity in $\alphabulge$ and $\alphadisc$
from galaxy to galaxy.
We plot the posterior PDF for the width parameters $\alphabulgewidth$ and
$\alphadiscwidth$ in \Fref{fig:hierarchical:widthalphas}. Note that the axes
in this plot are in logarithmic units, so that the assumed prior was uniform
inside the square shown.
The data have ruled out the upper and right-most parts of this space, and are
consistent with values in the lower left quadrant. A perfect universal IMF would
imply (among other things, such as $\alphabulgemedian = \alphadiscmedian$)
that $\alphabulgewidth = \alphadiscwidth = 0$, and this
is not included in our hypothesis space. However, low but nonzero values of
$\alphabulgewidth$ and $\alphadiscwidth$ may be considered close enough to zero
for practical purposes, but probability statements about universality would be
sensitive to the definition. Marginalising over all other hyperparameters and
parameters, we find 95\% upper limits by computing
$\pr(\alphabulgewidth \leq 0.165) = 95\%$
and $\pr(\alphadiscwidth \leq 0.375) = 95\%$. The prior probabilities of these
two hypotheses were 55\% and 64\% respectively.

The galaxy stellar mass fractions inferred in this hierarchical
model are plotted in Figure~\ref{fig:hierarchical:f}, where the error
bars represent the full uncertainty in the masses after marginalising
over $\alphabulge$  and $\alphadisc$ for that galaxy, and also over the
population hyper-parameters. 

To make this plot we repeated the inference while conditioning
on the Chabrier/Salpeter models by setting $\alphabulgemedian,\alphadiscmedian = (0.6,0.6), (0.6,1),
(1,1)$ and $(\alphabulgewidth, \alphadiscwidth) = (0, 0)$.
As opposed to Figure~\ref{fig:trend1} it can be immediately noted
that none of the inferred stellar mass fractions are unphysical, and
that all have smaller uncertainties than the simply-estimated ``observed''
mass fractions; this latter point results from the fact that our model
correlates each of the lenses and penalises widely discrepant masses,
thereby reducing the uncertainties.
In each case, we see a significant trend of lower stellar mass
fractions in galaxies with increasing total mass (as
represented by $\sigmasie$). This could be due to the dark matter
fraction increasing as the Einstein radius increases and the mass
enclosed becomes less dominated by the stars.


\section{Discussion and Conclusions}
\label{sec:discuss}
We have implemented a hierarchical model for a sample of strong lens
galaxies, and used it to interpret a set of highly compressed data:
total mass estimates obtained from strong gravitational lensing, and
stellar mass estimates of the bulge and disc components obtained from 
SPS models that assume a Salpeter IMF. This simple
set-up allowed us to explore the basic mechanics of the hierarchical
inference in some detail. We found that an apparently
uninformative prior on each galaxy component mass can actually have a
significant impact on the results, since this assignment opens up an
exponentially large volume of nuisance parameter space. A hierarchical
prior on the masses as well stabilises the problem, as well as matching
our actual knowledge of a) our selection process and b) galaxies in
general. Further work on higher fidelity data might profitably include
some relaxing of our assumptions of simple log-normal distributions for
both the component masses and the $\alpha$ parameters.

While the data set analysed here is only a small subset of all the available
information about variations of the IMF, it is an instructive subset that
demonstrates many of the issues that arise when combining small amounts of
information from large numbers of objects. In particular, the assumptions about
the prior distributions for nuisance parameters are extremely important:
assuming independence can lead to absurd conclusions.
We also note that scepticism of IMF variations \citep[e.g.][]{2010ARA&A..48..339B} 
must eventually break down at some
point as the only truly universal IMF would be one composed of delta function
spikes at the true initial mass of each actual star in the universe: in other
words, the appropriate question is not ``does the IMF vary'' but
``how much does the IMF vary, and as a function of what parameters?''.
Ultimately, detailed models of individual SWELLS galaxies, taking into account
all of the available data \citep{2012MNRAS.423.1073B} could be combined in
similar way to the analysis of this paper. This would lead to a full description
of the conclusions of the SWELLS survey: not only about the properties of the IMFs
of disk galaxies but also about all other properties measurable from gravitational
lensing and dynamical measurements.

While the analysis in this paper is something of a toy model, it
does achieve some important goals. First, we have replaced the rather
unsatisfactory ``observed stellar mass fractions'' that appeared in
SWELLS-III, and in the process, incorporated some new, quantitative
information -- the fact that stellar mass fractions cannot be greater
than one -- into the inference of IMF normalisation. Second, the
hierarchical framework we have presented has allowed us to explore  the
model that we were actually interested in! If we are to relax the
assumption of a Universal IMF, that means that each galaxy must have a
different IMF, each of which is to be measured. The hierarachical model
enables investigation of all possible scenarios in between the two
extremes of Universality and total randomness. It seems to us that
asserting some distribution of IMF parameters, implying some essential
similarity between galaxies in a sample, is the logical first step
beyond universality.


Our conclusions regarding the IMF normalisation in the bulge and disc
components of our sample of galaxies can be stated as follows:

\begin{itemize}
\item Under the assumptions that 1) the stellar populations of a particular
bulge or disc can each be described by a single IMF
normalisation, and 2) that bulges are drawn from one population
distribution (both in stellar mass and IMF normalisation) and discs
another, we find that the galaxies in the SWELLS sample seem to have
been drawn from a sub-population with lighter-than Salpeter IMF: the 
medians of the bulge and disc IMF normalisation distribution are such
that $p(\alphabulgemedian < 1) \approx 96.6\%$ while $p(\alphadiscmedian < 1)
\approx 81.6\%$.

\item The widths of the distributions of IMF normalisations are consistent
with small values (i.e. a close-to-universal IMF) or more moderate values
(variable IMF), both for SWELLS galaxy bulges and SWELLS
galaxy discs. Given the
uncertainty in the measurements, we cannot draw firm conclusions about
IMF universality from these parameters alone, despite them being the
natural choice. The results are consistent with the hypothesis of a universal
IMF, but not a universal Salpeter IMF.
\end{itemize}

We suggest that hierarchical modelling is the natural way to approach
inference problems involving samples of complex objects which can be
assumed to be similar in certain respects. In particular, as  samples of
strong lenses continue to grow, we expect these highly informative
individual  objects to be most informative in ensembles when analysed in
this way.

\section*{Acknowledgements}
It is a pleasure to thank Michael Williams (Columbia) for helpful
comments and discussions.
AAD acknowledges financial support from a CITA National Fellowship,
from the National Science Foundation Science and Technology Center
CfAO, managed by UC Santa Cruz under cooperative agreement
No. AST-9876783. AAD and DCK were partially supported by NSF grant AST
08-08133, and by HST grants AR-10664.01-A, HST AR-10965.02-A, and HST
GO-11206.02-A.
PJM was given support by the TABASGO and Kavli foundations, and the Royal 
Society, in the form of research fellowships.
TT acknowledges support from the NSF through CAREER award NSF-0642621,
and from the Packard Foundation through a Packard Research Fellowship.
LVEK acknowledges the support by an NWO-VIDI programme subsidy
(programme number 639.042.505).
This research is supported by NASA through Hubble Space Telescope
programs GO-10587, GO-10886, GO-10174, 10494, 10798, 11202, 11978,
12292 and in part by the National Science Foundation under Grant
No. PHY99-07949. and is based on observations made with the NASA/ESA
Hubble Space Telescope and obtained at the Space Telescope Science
Institute, which is operated by the Association of Universities for
Research in Astronomy, Inc., under NASA contract NAS 5-26555, and at
the W.M. Keck Observatory, which is operated as a scientific
partnership among the California Institute of Technology, the
University of California and the National Aeronautics and Space
Administration. The Observatory was made possible by the generous
financial support of the W.M. Keck Foundation. The authors wish to
recognize and acknowledge the very significant cultural role and
reverence that the summit of Mauna Kea has always had within the
indigenous Hawaiian community.  We are most fortunate to have the
opportunity to conduct observations from this mountain.
Funding for the SDSS and SDSS-II was provided by the Alfred P. Sloan
Foundation, the Participating Institutions, the National Science
Foundation, the U.S. Department of Energy, the National Aeronautics
and Space Administration, the Japanese Monbukagakusho, the Max Planck
Society, and the Higher Education Funding Council for England. The
SDSS was managed by the Astrophysical Research Consortium for the
Participating Institutions. The SDSS Web Site is http://www.sdss.org/.



\appendix
\onecolumn

\section{Probability Expressions}
Here we discuss the independence assumptions used in the
hierarchical model specified in \Sref{sec:model-hb}.
The unknown hyperparameters (which describe the
population distributions) are:
\begin{equation}
\boldsymbol{\eta} = \left\{
\alphabulgemedian, \alphabulgewidth,
\alphadiscmedian, \alphadiscwidth,
\mu_{\SPSMbulge}, \sigma_{\SPSMbulge},
\mu_{\SPSMdisc}, \sigma_{\SPSMdisc},
\ell_0, \ell_1, \sigma_\ell
\right\}
\end{equation}
The priors on these hyperparameters are given in \Tref{tab:everything}.
The parameters for an individual galaxy $j$ are:
\begin{equation}
\boldsymbol{\theta}^j =
\left\{\alphabulge^j, \alphadisc^j,
\SPSMbulge^j, \SPSMdisc^j, f^j\right\}.
\end{equation}
The joint prior for all unknown
hyperparameters {\it and} the individual galaxy parameters can be written
using the product rule:
\begin{eqnarray}
\pr\left(\boldsymbol{\eta}, \left\{\boldsymbol{\theta}^j\right\}_{j=1}^N\right)
&=& \pr\left(\boldsymbol{\eta}\right)
\pr\left(\left\{\boldsymbol{\theta}^j\right\}_{j=1}^N|\boldsymbol{\eta}\right)
\end{eqnarray}
The assumption is that, given the hyperparameters, the individual galaxy
parameters are independent:
\begin{eqnarray}
\pr\left(\boldsymbol{\eta}, \left\{\boldsymbol{\theta}^j\right\}_{j=1}^N\right)
&=& \pr\left(\boldsymbol{\eta}\right)
\prod_{i=1}^N
\pr\left(\boldsymbol{\theta}^j|\boldsymbol{\eta}\right)
\end{eqnarray}
where $\pr\left(\boldsymbol{\theta}^j|\boldsymbol{\eta}\right)$ is the same
function for each $j$. We also assume that the data for galaxy $j$ is independent
and only depends on the parameters for galaxy $j$:
\begin{eqnarray}
\pr\left(\mathbf{D}|\boldsymbol{\eta}, \boldsymbol{\theta}\right)
&=& \prod_{i=1}^N \pr\left(\mathbf{D}^j | \boldsymbol{\theta}^j\right)
\end{eqnarray}
where $\pr\left(\mathbf{D}^j | \boldsymbol{\theta}^j\right)$ is the same function
for all $j$.

One way to think about the hierarchical model is to think of the hyperparameters
as describing the distribution of galaxy properties in the population, and
each galaxy as being a point ``drawn from'' this distribution. The data is then
drawn independently (for each galaxy) from a distribution that is dependent
only on that galaxy's properties, and not on the properties of that galaxy or
the hyperparameters. These assumptions make sense when the objects being studied
can literally be thought of as having been sampled from some population.
However,
hierarchical
models can also be used when there is no population. Introducing the
hyperparameters
is a convenient way of constructing a prior distribution for $N$
{\it exchangeable}
variables that are dependent. One could imagine marginalising away
the
hyperparameters which would give a dependent prior on the parameters (such that,
for example, knowledge of several galaxies' properties would be informative
about the properties of a subsequent galaxy). hierarchical models can be used
to implement such dependent priors, even when the ``population'' motivation may
be absent.

\section{Corner/Triangle Plot}\label{sec:corner}
\begin{figure}
\begin{center}
\includegraphics[scale=0.4]{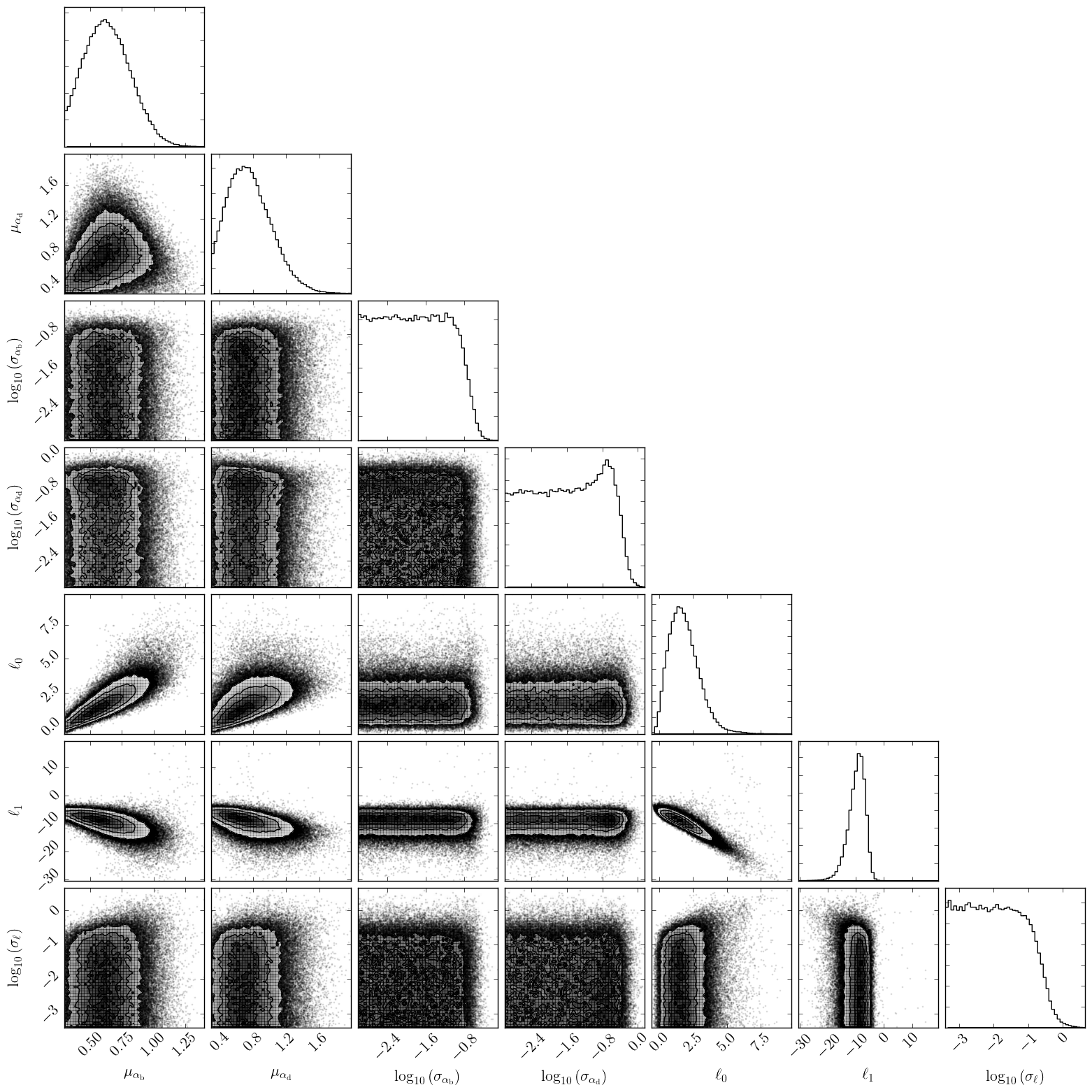}
\caption{A ``corner'' plot of joint posterior distributions for many of the
hyperparameters.\label{fig:corner}}
\end{center}
\end{figure}

For completeness, a ``corner'' plot, showing various joint posterior distributions
for an interesting subset of the hyperparameters, is given in Figure~\ref{fig:corner}. A distinct
correlation exists between $\ell_0$ and $\alphabulgemedian$, and $\ell_0$ and
$\alphadiscmedian$, reflecting the fact that higher stellar mass fractions
enable the IMF normalisation to be increased somewhat before becoming incompatible
with the data. This plot was made using {\tt triangle.py} by Dan Foreman-Mackey,
available from {\tt http://www.github.com/dfm/triangle.py}.

\section{Non-hierarchical Results}\label{sec:wrong}
\begin{figure}
\begin{center}
\includegraphics[scale=0.5]{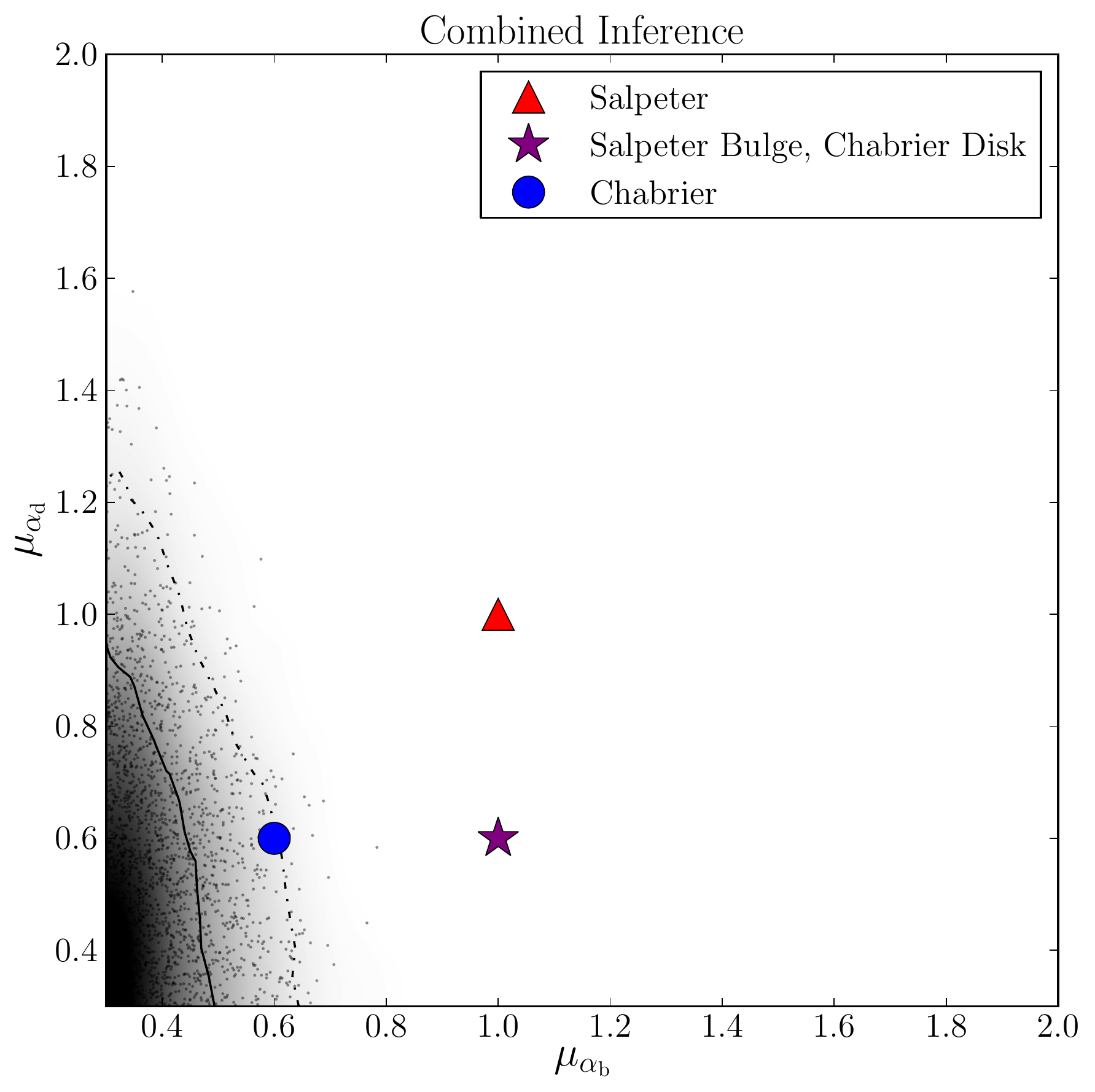}
\caption{The IMF inferences obtained by na\"{\i}ve application of the
single object model to all objects independently.\label{fig:wrong}}
\end{center}
\end{figure}

The results obtained by applying the single-object model
(Section~\ref{sec:model-onegalaxy}) to all 19 galaxies in the sample, and then
combining the results, is given in Figure~\ref{fig:wrong}. This model implies
strong evidence against all but the lightest IMF scenarios because it does not
recognise the possibility that a lot of the galaxies may have a high stellar
mass fraction (within the apertures used).

\label{lastpage}
\bsp

\end{document}